\documentstyle[12pt]{article}
\catcode`@=11
\newcount\@tempcntc
\def\@citex[#1]#2{\if@filesw\immediate\write\@auxout{\string\citation{#2}}\fi
  \@tempcnta\z@\@tempcntb\m@ne\def\@citea{}\@cite{\@for\@citeb:=#2\do
    {\@ifundefined
       {b@\@citeb}{\@citeo\@tempcntb\m@ne\@citea\def\@citea{,}{\bf ?}\@warning
       {Citation `\@citeb' on page \thepage \space undefined}}%
    {\setbox\z@\hbox{\global\@tempcntc0\csname b@\@citeb\endcsname\relax}%
     \ifnum\@tempcntc=\z@ \@citeo\@tempcntb\m@ne
       \@citea\def\@citea{,}\hbox{\csname b@\@citeb\endcsname}%
     \else
      \advance\@tempcntb\@ne
      \ifnum\@tempcntb=\@tempcntc
      \else\advance\@tempcntb\m@ne\@citeo
      \@tempcnta\@tempcntc\@tempcntb\@tempcntc\fi\fi}}\@citeo}{#1}}
\def\@citeo{\ifnum\@tempcnta>\@tempcntb\else\@citea\def\@citea{,}%
  \ifnum\@tempcnta=\@tempcntb\the\@tempcnta\else
   {\advance\@tempcnta\@ne\ifnum\@tempcnta=\@tempcntb \else \def\@citea{--}\fi
    \advance\@tempcnta\m@ne\the\@tempcnta\@citea\the\@tempcntb}\fi\fi}
\catcode`@=12
\def\barr{\begin{array}}
\def\earr{\end{array}}
\def\beq{\begin{equation}}
\def\eeq{\end{equation}}
\def\bea{\begin{eqnarray}}
\def\eea{\end{eqnarray}}
\def\bmath{\begin{displaymath}}
\def\emath{\end{displaymath}}
\def\bq{\begin{quote}}
\def\eq{\end{quote}}
\def\slash#1{\setbox0=\hbox{$#1$}#1\hskip-\wd0\hbox to\wd0{\hss\sl/\/\hss}}
\def\Slash#1{\mbox{$\not{\hspace{-1.03mm}#1}$}}

\def\as{\alpha_s}
\def\gE{\gamma_{\scriptscriptstyle E}}
\def\CF{C_{\scriptscriptstyle F}}
\def\Li{\mbox{$\mbox{\rm Li}_2$}}
\def\Frac#1#2{\mbox{$\textstyle{#1\over#2}$}}
\def\eps{\varepsilon}

\def\be{\begin{equation}}
\def\ee{\end{equation}}
\def\bea{\begin{eqnarray}}
\def\eea{\end{eqnarray}}
\def\nn{\nonumber\\}
\def\Slash#1{\mbox{$\not{\hspace{-1.03mm}#1}$}}

\def\as{\alpha_s}
\def\gE{\gamma_{\scriptscriptstyle E}}
\def\CF{C_{\scriptscriptstyle F}}
\def\Li{\mbox{$\mbox{\rm Li}_2$}}
\def\Frac#1#2{\mbox{$\textstyle{#1\over#2}$}}
\def\eps{\varepsilon}

\def\sw{s_{\scriptscriptstyle W}}
\def\cw{c_{\scriptscriptstyle W}}
\def\pint{\int{dy\,dz \over \sqrt{(1-y)^2-\xi}}\:\:}
\def\Frac#1#2{\mbox{$\textstyle{#1\over#2}$}}
\def\Li{\mbox{$\mbox{\rm Li}_2$}}
\def\uint{\int dy\,dz\:\:}
\def\real{\mathop{\rm Re}\nolimits}
\def\imag{\mathop{\rm Im}\nolimits}
\def\cz{\chi_{\scriptscriptstyle Z}}
\begin{document}
\begin{flushright}
MZ-TH/93-30 \\[-0.2cm]
RAL/93-81 \\[-0.2cm]
FTUV/93-46 \\[-0.2cm]
November 1993
\end{flushright}

\begin{center}
{\bf{\Large One-Loop {\em QCD} Mass Effects in the Production}}\\[0.3cm]
{\bf{\Large  of Polarized Bottom and Top Quarks}} \\[2.cm]
{\large J.G.~K\"orner}$^{\, a}$\footnote[1]{supported in part by the BMFT,
Germany under contract 06MZ730.},
{\large A.~Pilaftsis}$^{\, b}$\footnote[2]{e-mail address:
pilaftsis@vax2.rutherford.ac.uk}
{\large and M.M.~Tung}$^{\, c}$\footnote[3]{e-mail address:
tung@evalvx.ific.uv.es} \\[0.4cm]
$^{a}$ Institut f\"ur Physik, {\it THEP}, Johannes Gutenberg--Universit\"at,\\
55099 Mainz, Germany.\\[0.3cm]
$^{b}$ Rutherford Appleton Laboratory, Chilton, Didcot, Oxon, OX11 0QX,
England.\\[0.3cm]
$^{c}$ Departament de F\'\i sica Te\`orica, Univ.~de Val\`encia, and IFIC,\\
Univ.~de Val\`encia--CSIC, E-46100 Burjassot (Val\`encia), Spain.
\end{center}
\bigskip
\bigskip
\bigskip
\bigskip
\centerline {\bf ABSTRACT}

The analytic expressions for the production cross sections of polarized
bottom and top quarks in $e^+e^-$ annihilation are explicitly derived at the
one-loop order of strong interactions. Chirality-violating mass effects will
reduce the longitudinal spin polarization for the light quark pairs by an
amount of $3\%$, when one properly considers the massless limit for the final
quarks. Numerical estimates of longitudinal spin polarization effects in the
processes $e^+e^-\to b\bar{b}(g)$ and $e^+e^- \to t\bar{t}(g)$ are presented.
\newpage

\noindent
Precision tests at {\it LEP} have so far shown great agreement
with electroweak theory giving strong experimental support for the
Standard Model ({\it SM}\/)~\cite{GWS}. However, quantum chromodynamics
({\it QCD}\/),
though being non-convergent or weakly convergent in the perturbative expansion
at low energies, provides an interesting testing ground at the $Z$ peak to
probe many field-theoretical aspects of this asymptotically free theory.
In this note, we will study the longitudinal polarization asymmetry $P_{_L}$
of the bottom quark ($b$) and top quark ($t$)
produced through the $e^+e^-$ annihilation reaction at {\it LEP} and
future colliders. A suprising outcome of our calculations is that
the $O(\as )$ correction of $P_{_L}$ in the limit $m_q\to 0$
differs substancially from the corresponding result of a theory where $m_q$
was originally set to zero. In the following, we will call such a theory a
{\em naive massless theory}, since it leads to the wrong result
for the $O(\as)$ contributions to $P_{_L}$ for the process
$e^+e^-\to \gamma, Z \to b\bar{b}(g)$.

At the one-loop {\it QCD} level,
the $\gamma(q)-q(p_1)-\bar{q}(p_2)$ and $Z(q)-q(p_1)-\bar{q}(p_2)$
vertex functions relevant for the production of massive quarks can
be written down as follows:
\bea
\Gamma_\mu^\gamma &=&
-i e Q_q\,\left[\,(1+A)\gamma_\mu+B{(p_2-p_1)_\mu\over 2m_q}\,\right], \\
\Gamma_\mu^Z &=&
-i{e\over 4 \sw \cw}\left[\,
(1-4 \sw^2 Q_q)(1+A)\gamma_\mu
+(1-4\sw^2 Q_q)B{(p_2-p_1)_\mu\over 2m_q}\right. \nn
&&\left.-(1+C)\gamma_\mu\gamma_5
-D{(p_1+p_2)_\mu\over 2m_q}\gamma_5\,\right],
\eea 
where $\cw=M_W/M_Z$, $\sw=\sqrt{1-\cw^2}$ and
the form factors $A$, $B$, $C$ and $D$ have been calculated by using
dimensional regularization. These form factors are given by
\bea
A &=& {\as\over 4\pi}\CF\,\left[\,
\left({1+v^2\over v}\ln\left({1+v\over 1-v}\right)-2\right)
\left({2\over\eps}-\gE+\ln\left({4\pi\mu^2\over m_q^2}\right)\right) +
F(v)\,\right], \\
B &=& -{\as\over 4\pi}\CF\,{1-v^2\over v} \ln\left({1+v\over 1-v}\right), \\
C &=& {\as\over 4\pi}\CF\,\left[\,
\left({1+v^2\over v}\ln\left({1+v\over 1-v}\right)-2\right)
\left({2\over\eps}-\gE+\ln{4\pi\mu^2\over m_q^2}\right)\right.\nn
&&\left.+F(v)+2{1-v^2\over v}\ln\left({1+v\over 1-v}\right)\,\right], \\
D &=& -{\as\over 4\pi}\CF\,\left[\,(2+v^2){1-v^2\over v}
\ln\left({1+v\over 1-v}\right)-2(1-v^2)\,\right].
\eea 
The function $F(v)$ given in Eqs.~(3) and (5) is defined as
\bea
F(v) &=& \left[\,3v-{1+v^2\over 2v}\ln{4v^2\over 1-v^2}\,\right]
\ln\left({1+v\over 1-v}\right)\nn
& &+{1+v^2\over v}\left[\,\Li{v+1\over 2v}-\Li{v-1\over 2v}\,\right]
-4+{\pi^2\over 2}{1+v^2\over v},
\eea 
where $v=\sqrt{1-\xi}$ with $\xi=4m^2_q/q^2$.
To remove the {\it UV} divergences in the vertex functions
Eqs.~(1) and (2), we have considered the wave-function renormalization
constant of the final quarks
\be
Z_q = \left.
-{\partial\hphantom{\Slash{p}}\over\partial\Slash{p}}\Sigma_q(\Slash{p})\;
\right|_{\;\Slash{p}=m_q} = {\as\over 4\pi}\,\CF\left[
-{2\over\varepsilon_{\scriptscriptstyle UV}}+\gE-\ln{4\pi\mu^2\over m_q^2}
-4\right]
\ee 
and renormalized the form factors $A$ and $C$ according the prescription
\be
A=A_0+\delta Z_q,\qquad C=C_0+\delta Z_q\qquad.
\ee 
Our final result is also valid when employing dimensional reduction methods
for the calculation of $QCD$ quantum corrections~\cite{dimred}.
In addition to the $UV$ divergences, one encounters in Eqs.\ (3)--(6)
infrared ({\it IR}\/) singularities due to the soft-gluon part of the
one-loop contributions, which will exactly cancel with those of the real-gluon
emission graphs at the same order of strong interaction.

Decomposing the hadronic tensors in terms of
their Lorentz structure and considering the one-loop {\it QCD} corrections,
one gets
\bea
H^{VV}(virtual) &=& -q^2(3-v^2)-2A\,q^2(3-v^2)-2B\,q^2v^2, \\
H^{AA}(virtual) &=& -2q^2-4C\,q^2 v^2, \\
H^{VA}_\pm(virtual) &=& H^{AV}_\pm(virtual) = \mp (1+A+C) q^2 v,
\eea 
where the superscripts refer to the parity-parity combination of the
corresponding squared amplitudes.
A straightforward computation of the hadronic tensors involving
real-gluon emission gives
\bea
H^{VV}(real) &=&
-4{(1-y)^2+(1-z)^2\over y\,z}+2\xi\left(\,
{1\over y^2}+{1\over z^2}+{2\over y}+{2\over z}\,\right)+
\xi^2\left(\,{1\over y}+{1\over z}\,\right)^2, \\
H^{AA}(real) &=&
-4{(1-y)^2+(1-z)^2\over y\,z} -
2\xi^2 \left(\,{1\over y}+{1\over z}\,\right)^2 \nn
&&+\ 2\xi\left(\,
-2+{1\over y^2}+{1\over z^2}-{4\over y}-{4\over z}+{6\over yz}-{y\over z} -
{z\over y}\,\right), \\
\pm H^{VA}_\pm(real) &=& \pm H^{AV}_\pm(real) \nn
&=&
-\ 4(1-y){(1-y)^2+(1-z)^2\over y\,z\,\sqrt{(1-y)^2-\xi}}
+ {2\xi^2\over\sqrt{(1-y)^2-\xi}}\left[\,
\left({1\over y}+{1\over z}\right)^2-{2\over y^2}\,\right] \nn
&&+\ {2\xi\over\sqrt{(1-y)^2-\xi}}\left(\,
1+{1\over y^2}+{1\over z^2}-{5\over y}-{6\over z}+{6\over yz}+{y\over z}+
{z\over y}-{y\over z^2} \right. \nn
&&\left. +\ {y^2\over z^2}\,\right),
\eea 
where $y=1-2p_1\cdot q/q^2$ and $z=1-2p_2\cdot q/q^2$ are kinematical
variables.
Note that gluon-mass effects can safely be neglected
in the computation of the squared amplitudes, so that Eqs.~(13)--(15) only
display a quark-mass dependence through $\xi$.

The integration of Eqs.~(13)--(15) over the three-body phase-space
has been performed analytically. The various types of integrals contained
in the hadronic tensors are identified and listed in Appendix~A.
For compactness, it will be useful to re-express the hadronic tensors
in a different basis~\cite{KS}
\bea
H^1 &=& \Frac{1}{2}\left(\,H^{VV}+H^{AA}\,\right),   \\
H^2 &=& \Frac{1}{2}\left(\,H^{VV}-H^{AA}\,\right),   \\
H^3 &=& \Frac{i}{2}\left(\,H^{VA}-H^{AV}\,\right)=0, \\
H^4_\pm &=& \Frac{1}{2}\left(\,H^{VA}_\pm+H^{AV}_\pm\,\right)=H^{VA}_\pm.
\eea 
In this basis only $H^4$ depends on the spin orientation of the final
quark.
The production cross section of a polarized bottom or top quark can formally
be written as
\be
d\sigma_{tot} =
\sum_i\,g_i\,\bigl[\,H^i(virtual)\,dPS_2+H^i(real)\,dPS_3\,\bigr],
\ee 
where the electroweak couplings in the above basis, Eqs.\ (16)--(19), read
\bea
g_1 &=& Q_q^2-2 Q_q v_e v_q \real{\cz}+(v^2_e+a^2_e)(v_q^2+a_q^2) |\cz|^2, \\
g_2 &=& Q_q^2-2 Q_q v_e v_q \real{\cz}+(v^2_e+a^2_e)(v_q^2-a_q^2) |\cz|^2, \\
g_3 &=& -2 Q_q v_e a_q \imag{\cz}, \\
g_4 &=& 2 Q_q v_e a_q \real{\cz}-(v^2_e+a^2_e) 2 v_q a_q |\cz|^2.
\eea 
In Eqs.~(21)--(24) $Q_q$ denotes as usual the fractional charge of the
final-state quarks whereas $v_f=2T^f_z-4Q_f\sw^2$ and $a_f=2T^f_z$
are the electroweak
vector- and axial-vector coupling constants for fermions ($f$),
respectively. The couplings containing $\imag{\cz}$ and $\real{\cz}$ stem
from the $\gamma$--$Z$ interference. Here,
$\cz (s)= g_WM^2_Zs(s-M^2_Z+iM_Z\Gamma_Z)^{-1}$
characterizes the Breit-Wigner ({\it BW}\/) form of the $Z$ propagator, where
$M_Z$ and $\Gamma_Z$ are the mass and the total decay width of the $Z$ boson.
Note that we use a {\it BW} function with a constant decay width as
obtained by a Laurent series expansion in terms of the complex pole mass of
the $Z$ boson~\cite{RS}. This approach ensures the gauge invariance
in the vicinity of the $Z$-boson resonance~\cite{RS,AP}.
Denoting the two helicity states of the final quark $q$ by $\lambda_\pm=\pm
1/2$, one defines the following longitudinal polarization asymmetry:
\be
\left\langle\,P_{\scriptscriptstyle L}\,\right\rangle =
{\strut
 \sigma_{tot}\bigl(e^+ e^-\to q(\lambda_+)\bar{q}(g)\bigr) -
 \sigma_{tot}\bigl(e^+ e^-\to q(\lambda_-)\bar{q}(g)\bigr) \over\strut
 \sigma_{tot}\bigl(e^+ e^-\to q\bar{q}(g)\bigr) }.
\ee

At this point it is important to comment on the fact that in $O(\as)$
the mass-zero limit of the longitudinal polarization
does not coincide with the $O(\as)$ result of the naive massless theory
$\left\langle P_{\scriptscriptstyle L}\right\rangle_{m_q = 0}$.
In fact, one finds in the correct limit
\beq
\left\langle P_{\scriptscriptstyle L}\right\rangle_{m_q \to 0}\  =\
\frac{1+\alpha_s/3\pi}{1+\alpha_s/\pi}
\left\langle P_{\scriptscriptstyle L}\right\rangle_{m_q = 0}\ \simeq\ 0.975
\left\langle P_{\scriptscriptstyle L}\right\rangle_{m_q =0},
\eeq 
where it is
$\left\langle P_{\scriptscriptstyle L}\right\rangle_{m_q = 0} =-93.9\%$
for down-type quarks (i.e.\ $d$, $s$, $b$) and
$\left\langle P_{\scriptscriptstyle L}\right\rangle_{m_q = 0}= - 68.5\%$
for the up and charm quark at $q^2=M^2_Z$.
This suprising effect arises from the fact that there are
chirality-violating or helicity-flip contributions proportional to $m^2_q$
which are multiplied with a would-be singularity proportional to $1/m^2_q$.
Thus one has to include a finite spin-flip
contribution to $\left\langle P_{\scriptscriptstyle L}\right\rangle$
in the limit $m_q\to 0$, which evidently cannot be seen in the naive
massless theory. Similar observations have been reported by the authors
of Ref.~\cite{chiral}. In their analysis,
this chirality-violating effect originating solely from quantum
electromagnetic corrections was found to be small.
Here, this chirality-violating mass effect is of the order of $3\%$ (i.e.\
$2\as/3\pi$) and can, in principle, be observed at {\it LEP} by analyzing
$\left\langle P_{\scriptscriptstyle L}\right\rangle$ via the decay products
of the final quarks.

In Figs.~1 and~2 we have plotted the numerical values of
$\left\langle P_{\scriptscriptstyle L}\right\rangle$ versus the c.m.s.\
energy of the
$b$- and $t$-pair production, respectively. The solid line
is the Born approximation whereas the dashed line refers to the $O(\as)$
corrections. In particular, for the case of $b$-quark production the solid
line corresponds approximately to the line obtained in the naive massless
theory, while the dashed one represents the analytic result in the
limit $m_b \to 0$. A similar effect occurs in the top production which
is smaller of the order of $0.5\%$. For practical purposes, we present the
corresponding production cross sections for the $b$ and $t$ quark, which
have found to be in excellent agreement with Ref.~\cite{GR,KZ1,KZ2}.
For our numerical estimates we have used running masses and the running
coupling constant $\as$ with $\Lambda_{\overline{MS}}=0.238$~GeV~\cite{EAP}.

In conclusion, we have analytically calculated the longitudinal
polarization asymmetry of the final $b$ and $t$ quark for the processes
$e^+e^-\to b\bar{b} (g)$ and $e^+e^- \to t\bar{t} (g)$.
Especially, we have found that chirality-violating mass terms reduce
the spin-polarization asymmetry of the naive massless theory by
3$\%$ which is, in principle, quite sizable
to be detected in the {\it LEP} data. Since the top quark will
predominantly decay via electroweak interactions, spin polarization
asymmetry tests by analyzing the angular distribution of the charged leptons
arising from semileptonic top decays will furnish very attractive
and feasible experiments at future TeV $e^+e^-$ colliders.\\[2cm]
{\bf Acknowledgements.} We wish to thank J.~Bernab\'eu, J.~Smith, W.~Giele,
B.~Kniehl, M.~Lavelle and A.~Ilakovac for stimulating discussions.

\newpage

\setcounter{section}{1}
\setcounter{equation}{0}
\def\theequation{\Alph{section}\arabic{equation}}

\begin{appendix}

\section{Three-Body Phase Space}

\subsection*{I.~Spin-Independent Phase-Space Integrals}

The three body phase-space boundaries for the processes at hand
are given by
\bea
y_-&=& \Lambda^\frac{1}{2}\sqrt{\xi}+\Lambda\ , \qquad
y_+\ =\ 1-\sqrt{\xi} \\
z_\pm &=& \frac{2y}{\xi + 4y}\ \left(\
1-y-\frac{1}{2}\xi + \Lambda + \frac{\Lambda}{y}\ \pm\
\frac{1}{y}\sqrt{(1-y)^2-\xi} \sqrt{(y-\Lambda)^2 -\Lambda\xi}\ \right)
\eea
In Eqs.~(A1) and (A2) it is $\Lambda = \lambda^2/q^2$, where $\lambda$
denotes the gluon mass that has been introduced to regulate the soft
{\it IR} singularities.
After identifying the different types of integrals in Eqs.~(13) and (14) and
integrating over the above phase-space boundaries, one has
to the leading order of $\Lambda$~\cite{GR}
\bea
      I_1 & = &
      \uint \nn
      & = &
      \Frac{1}{2}v\left(1+\Frac{1}{2}\xi\right)-\Frac{1}{2}\xi\left(
      1-\Frac{1}{4}\xi\right)\ln\left({1+v\over 1-v}\right),\\
      I_2 & = &
      \uint {1\over y}\ =\ \uint {1\over z} \nn
      & = &
      -v+\left(1-\Frac{1}{2}\xi\right)\ln\left({1+v\over 1-v}\right),
\\
      I_3 & = &
      \uint {1\over y^2}\ =\ \uint {1\over z^2} \nn
      & = &
      -{4v\over\xi}\left(\ln\Lambda^{1\over 2}+\ln\xi-2\ln v-2\ln 2+1\right)+
      2\left(1-{3\over\xi}\right)\ln\left({1+v\over 1-v}\right),
\\
      I_4 & = &
      \uint {y\over z}\ =\ \uint {z\over y} \nn
      & = &
      -\Frac{1}{4}v\left(5-\Frac{1}{2}\xi\right)+\Frac{1}{2}\left(
      1+\Frac{1}{8}\xi^2\right)\ln\left({1+v\over 1-v}\right),
\\
      I_5 & = &
      \uint {1\over y z} \nn
      & = &
      \left(-2\ln\Lambda^{1\over 2}-\ln\xi+4\ln v+2\ln 2\right)
      \ln\left({1+v\over 1-v}\right) \nn
      & &
     +2\left[\Li\left({1+v\over 2}\right)-\Li\left({1-v\over 2}\right)\right]
   +3\left[\Li\left(-{2v\over 1-v}\right)-\Li\left({2v\over 1+v}\right)\right],
\\
      I_6 & = &
      \uint {y\over z^2}\ =\ \uint {z\over y^2} \nn
      & = &
      {2\over\xi}\,v + \ln\left({1-v\over 1+v}\right).
\eea
Note that the expressions $I_1-I_5$ agree with the ones given in~\cite{GR}
whereas we get a different result for $I_6$.
\vspace{1cm}
\subsection*{II.~Spin-Dependent Phase-Space Integrals}

To the best of our knowledge, the following analytic expressions for
the basic set of phase-space integrals involving spin-dependent
integrand functions have not been
presented in the literature before.
Due to the presence of the additional square root in the integrand and
the rather involved phase-space boundaries a number of sophisticated
manipulations had to be applied to arrive at analytic expressions.
After straightforward but tedious algebra and neglecting all terms
that vanish in the limit $\Lambda\to 0$, we obtain the following expressions:
\bea
      S_1 & = &
      \pint \nn
      & = &
      1-\sqrt{\xi}-\Frac{1}{2}\xi\ln\left({2-\sqrt{\xi}\over\sqrt{\xi}}\right),
\\
      S_2 & = &
      \pint {1\over y} \nn
      & = &
      2 \ln\left({2-\sqrt{\xi}\over\sqrt{\xi}}\right),
\\
      S_3 & = &
      \pint {1\over y^2} \nn
      & = &
      {4\over\xi} \Biggl[ -\ln\Lambda^{1\over 2}+\Frac{1}{2}\ln\xi+
      \ln(1-\sqrt{\xi})-2\ln(2-\sqrt{\xi})+\ln 2-1 \Biggr],
\\
      S_4 & = &
      \pint {1\over z} \nn
      & = &
      \Li\left({1+v\over 2}\right)+\Li\left({1-v\over 2}\right)+
      2\,\Li\left(-{\sqrt{\xi}\over 2-\sqrt{\xi}}\right)+\Frac{1}{4}
      \ln^2\xi \nn
      & &
      +\ln^2\left({2-\sqrt{\xi}\over 2}\right)-\ln(1+v)\ln(1-v),
\eea
\newpage
\bea
      S_5 & = &
      \pint {1\over z^2} \nn
      & = &
      {4\over\xi} \left[ -\ln\Lambda^{1\over 2}-\Frac{1}{2}\ln\xi+
      \ln(1-\sqrt{\xi})-{1+v^2\over 2v}
      \ln\left({1+v\over 1-v}\right)+\ln 2 \right],
\\
      S_6 & = &
      \pint {y\over z^2} \nn
      & = &
      {4\over\xi} \left(1-\sqrt{\xi}\right),
\\
      S_7 & = &
      \pint {y^2\over z^2} \nn
      & = &
      {2\over\xi} \left(1-\sqrt{\xi}\right)^2,
\\
      S_8 & = &
      \pint z \nn
      & = &
      {1\over 32} \left[ 12-(2+\xi)^2-{2+\sqrt{\xi}\over 2-\sqrt{\xi}}\,\xi^2
      +2(8-\xi)\xi\,\ln{\sqrt{\xi}\over 2-\sqrt{\xi}} \right],
\\
      S_9 & = &
      \pint {z\over y} \nn
      & = &
      -\Frac{1}{2}\ln\xi+\ln\left(2-\sqrt{\xi}\right)+
      {2\over 2-\sqrt{\xi}}-2,
\\
      S_{10} & = &
      \pint {y\over z} \nn
      & = &
      \Li\left({1+v\over 2}\right)+\Li\left({1-v\over 2}\right)-
      2\,\Li\left({\sqrt{\xi}\over 2}\right)+\Frac{1}{4}
      \ln^2\bigl(\Frac{1}{4}\xi\bigr) \nn
      & &
      +\left(2-\Frac{1}{2}\xi\right)
      \ln\left({2-\sqrt{\xi}\over\sqrt{\xi}}\right)-\sqrt{\xi}+
      2v\ln\left({1-v\over 1+v}\right) \nn
      & &
      -\ln\left({1+v\over 2}\right)\ln\left({1-v\over 2}\right)+1,
\\
      S_{11} & = &
      \pint {y^2\over z} \nn
      & = &
      \left(1+\Frac{1}{2}\xi\right) \left[ \Li\left({1+v\over 2}\right)+
      \Li\left({1-v\over 2}\right)-2\,\Li\left(\Frac{1}{2}\sqrt{\xi}\right)+
      \Frac{1}{4}\ln^2\left(\Frac{1}{4}\xi\right) \right. \nn
      & &
      \left. -\ln\left({1+v\over 2}\right)\ln\left({1-v\over 2}\right) \right]+
      3v \ln\left({1-v\over 1+v}\right)+
      \Frac{1}{8}(18+\xi)-\Frac{1}{8}(20-\xi)\sqrt{\xi} \nn
      & &
      +\left(3-\xi+\Frac{1}{16}
      \xi^2\right) \ln\left({2-\sqrt{\xi}\over\sqrt{\xi}}\right),
\eea
\newpage
\bea
      S_{12} & = &
      \pint {1\over y z} \nn
      & & \nn
      & = &
      {1\over v}\ln\left({1-v\over 1+v}\right) \left[ 2\ln\Lambda^{1\over 2} +
      \Frac{1}{2}\ln\xi+4\ln(2-\sqrt{\xi})-4\ln v-4\ln 2-
      2\ln\left({1-v\over 1+v}\right) \right] \nn
      & & \nn
      & &
      + {1\over v}\ln^2\left({(1-v)^2\over\sqrt{\xi}(2-\sqrt{\xi})}\right)+
      {2\over v} \ln\left({\sqrt{\xi}(2-\sqrt{\xi})\over 2}\right)
      \ln\left({2\sqrt{\xi}(1-\sqrt{\xi})\over (1-\sqrt{\xi}-v)^2}\right)
      \nn
      & & \nn
      & &
      +{2\over v} \left[ \Li\left({\sqrt{\xi}(2-\sqrt{\xi})\over (1+v)^2}
      \right)-\Li\left[\left({1-v\over 1+v}\right)^2\right]+
      \Li\left({(1-v)^2\over\sqrt{\xi}(2-\sqrt{\xi})}\right) \right]
      \nn
      & & \nn
      & &
      +{1\over v} \left[ \Li\left({1+v\over 2}\right)-
      \Li\left({1-v\over 2}\right)+\Li\left(-{2v\over 1-v}\right)-
      \Li\left({2v\over 1+v}\right)-{\pi^2\over 3} \right].
\eea
\end{appendix}

\newpage

\centerline{\bf\Large Figure Captions }
\vspace{1cm}
\newcounter{fig}
\begin{list}{\bf\rm Fig. \arabic{fig}: }{\usecounter{fig}
\labelwidth1.6cm \leftmargin2.5cm \labelsep0.4cm \itemsep0ex plus0.2ex }

\item Longitudinal polarization asymmetry for bottom quark as a function
of the c.m.s.~energy. The solid line corresponds to the Born approximation
(or the $O(\as)$ result of the naive massless theory) while the dashed one
is calculated at the one-loop $QCD$ order. We have used $\alpha_s(M^2_Z)=0.12$.

\item Longitudinal polarization asymmetries for three different masses
of top quark as a function of the c.m.s.\ energy. The dashed and dotted lines
are indicated similar to Fig.~1.

\item {\bf (a).} Production cross section for the bottom quark as a function
of the c.m.s.~energy. {\bf (b).} Production cross section of the
$b$ quark in the vicinity of the $Z$-boson mass,
where the solid line represents
the born approximation, the dotted one denotes the one-loop $QCD$ result
for $m_b\to 0$ and the dashed line gives the massive one-loop $O(\as)$
corrections.

\item Production cross sections for three different masses of top quark.

\end{list}


\newpage
\begin{thebibliography}{99}

\bibitem{GWS} S.L.~Glashow, Nucl.\ Phys.\ {\bf 22} (1961) 579;
S.\ Weinberg, Phys.\ Rev.\ Lett.\ {\bf 19} (1967) 1264;
A.\ Salam, Proc.\ of 8th Nobel Symposium 1968, ed. N.\ Svarthom,
Wiley 1968, p.~367.

\bibitem{dimred} J.G.~K\"orner and M.M.~Tung, preprint MZ-TH/92-41

\bibitem{KS} J.G.~K\"orner and D.H.~Schiller, preprint DESY-81-043
(unpublished);
H.A.~Olsen, P.~Osland and I.~\O verb\o ,
Nucl.\ Phys.\ {\bf B171} (1980) 209.

\bibitem{RS} R.G.~Stuart, Phys.\ Lett.\ {\bf B262} (1991) 113;
A.\ Sirlin, Phys.\ Rev.\ Lett.\ {\bf 67} (1991) 2127.

\bibitem{AP} M.~Nowakowski and  A.~Pilaftsis, Z.\ Phys.\ {\bf C60}
(1993) 121; and references therein.

\bibitem{chiral} Z.~Was, Acta Phys.\ Pol.\ {\bf B18} (1987) 1099;
A.V.~Smilga, Comm.\ Nucl.\ Part.\ Phys.\ {\bf 20} (1991) 69;
B.~Falk and L.M.~Sehgal, Aachen preprint (1993), PITHIA 93/29.

\bibitem{GR} G.~Grunberg, Y.J.~Ng and S.-H.H.\ Tye, Phys.\ Rev.\
{\bf D21} (1980) 62.

\bibitem{KZ1} J.H.~K\"uhn, A.~Reiter and P.M.~Zerwas, Nucl.\ Phys.\
{\bf B272} (1986) 560; A.~Djouadi, Z.\ Phys.\ {\bf C39} (1988) 561.

\bibitem{KZ2} A.~Djouadi, J.H.~K\"uhn and P.M.~Zerwas, Z.\ Phys.\
{\bf C46} (1990) 411; P.M.~Zerwas, preprint DESY-93-001;
E.~Laenen, J.~Smith and W.L.~van Neerven, Fermilab preprint (1993),
FERMILAB-Pub-93/270-T.

\bibitem{EAP} See for example K.\ Kang, J.\ Flanz, E.A.\ Paschos,
Z.\ Phys.\ {\bf C55} (1992) 75.

\end{thebibliography}
\end{document}